\documentclass[11pt]{article}
\usepackage{times}
\usepackage{amsmath,amssymb,amsfonts,amsthm,mathrsfs}
\usepackage{graphicx}
\usepackage[colorlinks=true,urlcolor=blue]{hyperref}
\def\be{\begin{equation}}
\def\ee{\end{equation}}
\def\bea{\begin{eqnarray}}
\def\eea{\end{eqnarray}}
\def\lb{\label}
\def\ct{\cite}
\def\bi{\bibitem}
\begin{document}
\title{\sc Decision-theoretic approaches to non-knowledge in 
economics}
\author{
{\sc Ekaterina Svetlova}${}^{1,2}$\thanks{E--mail:
{\tt esvetlova@yahoo.de}}
\ and \ 
{\sc Henk van Elst}${}^{2}$\thanks{E--mail:
{\tt hvanelst@karlshochschule.de}} \\
{\small ${}^{1}$\emph{The Center of Excellence ``Cultural 
Foundations of Integration,'' Institute for Advanced Studies}} \\ 
{\small\emph{Universit\"{a}t Konstanz, Otto--Adam--Stra\ss e 5, 
78467 Konstanz, Germany}} \\
{\small ${}^{2}$\emph{Fakult\"{a}t I: Betriebswirtschaft und
Management}} \\
{\small\emph{Karlshochschule International University,
Karlstra\ss e 36--38, 76133 Karlsruhe, Germany}}}

\date{\normalsize{July 3, 2014}}
\maketitle
\begin{abstract}
We review two strands of conceptual approaches to 
the formal representation of a decision maker's non-knowledge at 
the initial stage of a static one-person, one-shot decision 
problem in economic theory. One focuses on representations of 
non-knowledge in terms of probability measures over sets of 
mutually exclusive and exhaustive consequence-relevant states of 
Nature, the other deals with unawareness of potentially important 
events by means of sets of states that are less complete than the 
full set of consequence-relevant states of Nature. We supplement 
our review with a brief discussion of unresolved matters in both 
approaches.

\medskip
\noindent
To appear in \emph{Routledge International Handbook of Ignorance 
Studies}, edited by Matthias Gro\ss\ and Linsey McGoey (London: 
Routledge), due to be published in February 2015.

\end{abstract}

\section{Introduction}
\lb{sec1}
The aim of this contribution is to provide an overview of conceptual approaches to incorporating a decision maker's non-knowledge into economic theory. We will focus here on the particular kind of non-knowledge which we consider to be one of the most important for economic discussions: non-knowledge of possible consequence-relevant uncertain events which a decision maker would have to take into account when selecting between different strategies.

\medskip
\noindent
It should be noted that --- especially after the recent worldwide economic crisis --- economics has been frequently blamed for neglecting this kind of non-knowledge. Allegedly it failed to incorporate unexpected events into its theoretical framework, which resulted in severe negative consequences for economies and societies. (For example, the subprime mortgage crisis or the Lehman Brothers bankruptcy of 2008 can be viewed as ``Black Swan events'' in the sense of Taleb (2007)~\ct{tal2007}). We argue, however, that such blatant accusations are not entirely justified. When one looks back at the long history of the debate on uncertainty and non-knowledge in economics, one will identify ongoing efforts to formalize these conceptually difficult issues by means of the mathematical language on the one hand, and tireless criticisms of this formal approach on the other. The first movement is often interpreted as essentially excluding non-knowledge from economic theory, while the second is considered as a heroic effort to re-establish this issue in the scientific discourse (Frydman and Goldberg 2007~\ct{frygol2007}; Akerlof und Shiller 2009~\ct{akeshi2009})). However, we would like to stress and demonstrate that both developments are deeply interwoven and, rather, mutually support and complement each other.

\medskip
\noindent
In the course of the debate, the theoretical representations of non-knowledge have taken some specific technical forms. In this paper, we review the historical development of two basic approaches to formalizing non-knowledge in economic theory, in the context of static one-shot choice situations for decision makers. These are
\begin{enumerate}
\item representations of non-knowledge of a decision maker in terms of \emph{probability measures, or related non-probabilistic measures}, over sets of mutually exclusive and exhaustive consequence-relevant (past, present or, in most applications, future) states of Nature;

\item modelling unawareness of a decision maker of potentially important events by means of sets of states that are \emph{less complete} than the full set of consequence-relevant states of Nature.
\end{enumerate}

\medskip
\noindent
As is well known, the most popular method to deal with non-knowledge in economic theory has been to formalize it by means of probability measures; this approach allowed quantifying the matter and, thus, to rationalize and to ``cultivate'' it (Smithson 1989~\ct[p~43]{smi1989}). Introduced into economic theory by Edgeworth, Jevons and Menger during the so called ``marginal revolution'' in the late 19\emph{th} century, probability measures, especially frequentist probability measures as probabilities learned from the past, were celebrated as instruments that allowed quantifying and measuring manifestations of uncertainty (cf. Bernstein 1996~\ct[p~190ff]{ber1996}). However, the euphoria was halted by the critiques of Knight (1921)~\ct{kni1921}, Keynes (1921, 1937)~\ct{key1921,key1937}, Shackle (1949, 1959)~\ct{sha1949,sha1959} and Hayek (1945)~\ct{hay1945} who argued that application of frequentist probability measures precludes systematic analysis of the principal non-knowledge of some consequence-relevant events. They initiated the first line of discussion on non-knowledge in economics and decision-making theory; namely, they raised the question as to what extent non-knowledge can be represented by means of measurable or immeasurable probability concepts, or if other, non-probabilistic measures are necessary. Knight's (1921)~\ct{kni1921} solution, for example, was the famous distinction between \emph{risk} as situations where probabilities of uncertain events can be unambiguously and objectively determined, and \emph{uncertainty} as situations where they cannot be accurately measured and, therefore, should rather be treated as ``estimates of the estimates,'' or subjective probabilities.

\medskip
\noindent
This critique gave rise to an axiomatic approach to the definition of subjective probability measures by Ramsey (1931)~\ct{ram1931} and de Finetti (1937)~\ct{fin1937} who demonstrated that such measures can always be derived from the observed betting behaviour of a decision maker (namely their willingness to bet), and that they can be powerfully used to formalize a decision maker's proclaimed utility maximization. Both authors helped to establish the concept of \emph{probabilistic sophistication} which posits that --- even if objective probability measures cannot be determined --- the decision maker's behaviour can always be interpreted in a way \emph{as if} they have a subjective probability measure which they employ in their personal calculations of expected utility. In this approach, \emph{individual} imprecise knowledge on consequence-relevant events was conceptualized to form a basis for the introduction of an adequate probability measure to represent this status, and this method rendered the whole discussion about measurability and objectivity of probability measures obsolete for the coming years. Savage (1954)~\ct{sav1954} famously combined probabilistic sophistication with the expected utility theory as conceived originally by Bernoulli (1738)~\ct{ber1738} and von Neumann and Morgenstern (1944)~\ct{neumor1944} to arrive at a subjective expected utility theory. Savage's axiomatization of decision-making under conditions of uncertainty thus led to the formalization of non-knowledge of the likelihood of uncertain events in terms of a \emph{unique (finitely additive) Bayes--Laplace prior probability measure} over a complete space of consequence-relevant states of Nature. The latter is assumed to be known to a decision maker before committing to a certain action.

\medskip
\noindent
Yet, this theoretical move to ``absorb'' non-knowledge by means of probability distributions obviously precludes the consideration of ``unknown unknowns'' (e.g. Li 2009~\ct[p~977]{li2009}), as, by assumption, this space of states of Nature is common knowledge for all decision makers. The prior probability measures employed just formalize non-knowledge of \emph{which} uncertain event from a given list of possibilities will occur. The incorporation of \emph{surprises} into a theoretical framework, however, necessitates a notion of incomplete sets of uncertain events on the decision maker's part. Surprising events, by definition, cannot be known at the instant of choice and, thus, cannot be part of the set of events possible known to a decision maker. However, many accounts which aspire to introduce true non-knowledge and uncertainty into economic theory primarily criticize the use of a unique and additive prior probability measure over a given set of states of Nature, but maintain the assumption that the latter set is finite and exhaustive, and that the states are mutually exclusive. These works thus pursue the first line of research mentioned above. In their attempt to formally deal with true uncertainty of some events, and so to re-establish this issue in economic theory, they replace the unique, additive prior probability measure by entire \emph{sets} of additive prior probability measures (e.g. Gilboa and Schmeidler 1989~\ct{gilsch1989}, Bewley 1986, 2002~\ct{bew1986,bew2002}), by non-additive prior probability measures (e.g. Schmeidler 1989~\ct{sch1989}, Mukerji 1997~\ct{muk1997}, Ghirardato 2001~\ct{ghi2001}), or they introduce some alternative non-probabilistic concept such as fuzzy logic, possibility measures, and weights (Zadeh 1965, 1978~\ct{zad1965,zad1978}; Dubois and Prade 2011~\ct{dubpra2011}; Kahneman and Tversky 1979~\ct{kahtsv1979}).

\medskip
\noindent
We interpret all of these works as attempts to conceptualize Knightian uncertainty in mathematical terms. In all of these cases, non-knowledge is generally captured by an \emph{unknown probability measure}. However, we would like to stress that theorizing about the principal non-knowledge of some events necessitates the aforementioned representation of the \emph{incompleteness of a decision maker's subjective space of consequence-relevant states of Nature}, because only then the failure of probability theory to represent non-knowledge and surprises adequately can be overcome. This state of affairs motivates the discussion of the second line of research on non-knowledge in the list above. Here, we are dealing with attempts to formalize choice situations where decision makers are aware of the fact that they do \emph{not} possess the full list of consequence-relevant states of Nature due to unforeseen contingencies. In our view, the development of this second line shifts emphasis from the issue of the importance of prior probability measures in dealing with non-knowledge to the more fundamental question as to what extent the full space of consequence-relevant states of Nature can actually be known to a decision maker in the first place.

\medskip
\noindent
In what follows, we first present in Section~\ref{sec2} the standard mathematical framework in terms of which discussions on the formal representation of non-knowledge and uncertainty in economic theory are usually conducted. Subsequently, we describe developments of the inclusion/exclusion movements of non-knowledge along the two lines mentioned above. In Section~\ref{sec3} we address the representation of non-knowledge based on the usage of (various kinds of) probability measures, while in Section~\ref{sec4} we discuss the representation of non-knowledge based on particular formal descriptions of the state space. In Section~\ref{sec5} we conclude with a discussion and provide a brief outlook.

\section{The basic mathematical framework}
\lb{sec2}
In economic theory, non-knowledge of the likelihood of uncertain events at the initial decision stage of a static one-person, one-shot decision problem is the crucial feature. Formulated within the set-theoretic descriptive behavioural framework developed by Savage (1954)~\ct{sav1954} and Anscombe and Aumann (1963)~\ct{ansaum1963}, a decision maker chooses from a set $\boldsymbol{F}$ of alternative acts. The set $\boldsymbol{\Delta}(\boldsymbol{X})$ of consequences of their choice (i.e., von Neumann--Morgenstern (1944)~\ct{neumor1944} lotteries over sets $\boldsymbol{X}$ of outcomes) depends on which relevant state of Nature out of an exclusive and exhaustive set $\boldsymbol{\Omega}$ will occur following a decision (the state-contingency structure). In this framework, acts are perceived as mappings of states of Nature into consequences, $\boldsymbol{F}=\{f:\boldsymbol{\Omega}\rightarrow\boldsymbol{\Delta}(\boldsymbol{X})\}$. An ordinal binary preference relation $\succeq$ is defined over the set $\boldsymbol{F}$, which in turn induces an analogous preference relation on the set $\boldsymbol{\Delta}(\boldsymbol{X})$ via the mapping. The actual state of Nature $\omega \in \boldsymbol{\Omega}$ that \emph{will} be realized is usually understood as a move by the exogenous world which resolves all uncertainty (Nature ``chooses'' the state of the world; Debreu 1959~\ct{deb1959}, Hirshleifer and Riley 1992~\ct[p~7]{hirril1992}). The decision maker \emph{does not know} which consequence-relevant state of Nature will occur, but (like the modeller) has complete knowledge of all possibilities. In the subjective expected utility context of Savage (1954)~\ct{sav1954} and Anscombe and Aumann (1963)~\ct{ansaum1963}, this kind of non-knowledge is formalized by means of a unique finitely additive prior probability measure over the set of states of Nature, $\mu \in \boldsymbol{\Delta}(\boldsymbol{\Omega})$, which expresses the decision maker's assessment of the likelihood of all uncertain events possible. Existence of such a prior probability measure (usually interpreted as representing a decision maker's beliefs) is ensured provided the preference relation $\succeq$ on $\boldsymbol{F}$ satisfies the five behavioural axioms of weak order, continuity, independence, monotonicity and non-triviality (see Anscombe and Aumann 1963~\ct[p~203f]{ansaum1963}, Gilboa 2009~\ct[p~143f]{gil2009}). As is well known, this axiomatization gives rise to a subjective expected utility representation of the preference relation $\succeq$ on $\boldsymbol{F}$ in terms of a real-valued preference function $V:\boldsymbol{F} \rightarrow \mathbb{R}$, where for every act $f \in \boldsymbol{F}$ one defines
\[ V(f)=\int_{\boldsymbol{\Omega}}\left(E_{f(\omega)}U\right)\mu({\rm d}\omega) \ .
\]
Here $U:\boldsymbol{X} \rightarrow \mathbb{R}$ constitutes a decision maker's real-valued personal utility function of an outcome $x \in \boldsymbol{X}$, and it is unique up to positive linear transformations; $E_{f(\omega)}U$ denotes the expectation value of $U$ with respect to the von Neumann--Morgenstern lottery $f(\omega) \in \boldsymbol{\Delta}(\boldsymbol{X})$. The decision maker weakly prefers an act $f \in \boldsymbol{F}$ to an act $g \in \boldsymbol{F}$, namely $f \succeq g$, whenever $V(f) \geq V(g)$. It is presupposed here that the prior probability measure is used to express the non-knowledge of the decision maker about exactly \emph{which state (from the given exhaustive list) will occur}. Figure~\ref{fig:fig1} outlines the structure of the decision matrix for a static one-person, one-shot choice problem in the subjective expected utility framework due to Savage (1954)~\ct{sav1954} and Anscombe and Aumann (1963)~\ct{ansaum1963}. We remark in passing that the exposition by the latter two authors in particular provided the formal basis for more recent decision-theoretical developments by Schmeidler (1989)~\ct{sch1989}, Gilboa and Schmeidler (1989)~\ct{gilsch1989} and Schipper (2013)~\ct{sch2013}.
\begin{figure}[!htb]
\begin{center}
\[
\begin{array}{c|cccc|c}
\text{prior probability measure}\ \mu\in\boldsymbol{\Delta}(\boldsymbol{\Omega}) & \mu(\omega_{1}) & \mu(\omega_{2}) & 
\ldots & \mu(\omega_{n}) & n \in \mathbb{N} \\
\hline
\text{acts}\ \boldsymbol{F}\ \backslash\ \text{states}
\ \boldsymbol{\Omega} & \omega_{1} & 
\omega_{2} & \ldots & \omega_{n} & \\
\hline
f_{1} & p_{11} & p_{12} & \ldots & p_{1n} & \text{consequences}
\ p_{ij} \in \boldsymbol{\Delta}(\boldsymbol{X}) \\
f_{2} & p_{21} & p_{22} & \ldots & p_{2n} & (\text{lotteries over 
outcomes} \ \boldsymbol{X}) \\
\vdots & \vdots & \vdots & \ddots & \vdots & 
\end{array} \ ,
\]
\end{center}
\caption{One-person, one-shot decision matrix of static decision 
problems in subjective expected utility theory.}
\lb{fig:fig1}
\end{figure}

\medskip
\noindent
The decision matrix suggests that, besides coding uncertainty about the likelihood of events via a unique prior probability measure, there are at least two further ways to incorporate aspects of a decision maker's non-knowledge: \emph{either} to suppose that the particular kind of prior probability measure is unknown (while the set of consequence-relevant states of Nature is complete), \emph{or} to accept that the set of consequence-relevant states of Nature can be known only incompletely. In the second case, non-knowledge of events is directly captured by means of \emph{non-knowledge of the full state space, allowing hereby for unexpected events}. In what follows, we will discuss these two general possibilities to formally deal with non-knowledge in more detail --- the application of various probability measures on the one hand, and the representation of incomplete state spaces on the other.

\section[Probabilistic and non-probabilistic approaches]{First way 
of formalization: probabilistic and non-probabilistic approaches}
\lb{sec3}
As already mentioned, application of additive prior probability measures to capture non-knowledge about the likelihood of uncertain events has been the silver bullet of economic theory in dealing with this problem. In terms of elements of the decision matrix in Figure~\ref{fig:fig1}, \emph{both} the modeller and the decision maker have complete knowledge of all consequence-relevant states of Nature, and of all possible outcomes/lotteries over outcomes contingent on these states. In this respect, both subjects need to be perceived as omniscient. However, throughout the entire history of applications of probability theory in various manifestations in the economic science, discussions have revolved around the question whether different kinds of definitions of probability measures are measurable at all in economic settings, and thus suitable to represent non-knowledge of some events. According to Knight (1921)~\ct{kni1921}, the conceptual basis for such an operationalization is principally absent from economic life in most cases. Thus, mathematical and statistical probabilities are --- though basically measurable --- not applicable in an economic context.

\medskip
\noindent
The concerns as formulated in the works of Knight (1921)~\ct{kni1921}, Keynes (1921)~\ct{key1921}, and also Shackle (1949)~\ct{sha1949}, however, were played down for a while by the opposing movement of strong formalization and specific exclusion of \emph{non-knowledge of the probability measure} from the theoretical economic framework: Ramsey (1931)~\ct{ram1931}, de Finetti (1937)~\ct{fin1937} and Savage (1954)~\ct{sav1954} demonstrated that subjective probabilities can be measured in principle when taking a behavioural approach. Follow-up research, however, drew attention to cases in which non-knowledge of the probability measure is essential for decision-making. Especially after Ellsberg's (1961)~\ct{ell1961} paper, a new branch of research appeared that endeavoured to re-introduce absence of perfect knowledge of relevant probability measures into economic theory. Ellsberg (1961)~\ct{ell1961} had demonstrated empirically that many people tend to prefer situations with known probability measures over situations with unknown ones, thus violating Savage's (1954)~\ct{sav1954} behavioural ``sure thing principle'' axiom in particular. He explicitly referred to situations with unknown probability measures as ``ambiguous'' and named the phenomenon of avoiding such situations ``ambiguity aversion'' (this corresponds to the term ``uncertainty aversion'' coined by Knight 1921~\ct{kni1921}).

\medskip
\noindent
Subsequently, efforts to formalize \emph{Knightian uncertainty} were resumed. Relevant work has been developing in two directions (cf. Mukerji 1997~\ct[p~24]{muk1997}). First, it was stressed that, in Savage's (1954)~\ct{sav1954} static choice framework, the decision maker `mechanically' assigns probabilities without differentiating between those cases in which they have some knowledge and, thus, can reason about the likelihood of future events, and those cases in which they are completely ignorant about what might happen. Secondly, Savage's (1954)~\ct{sav1954} framework precludes from modelling the decision maker ``\ldots who doubts his own ability to imagine and think through an exhaustive list of possible states of the world'' (Mukerji 1997~\ct[p~24]{muk1997}). Savage's (1954)~\ct{sav1954} axiomatization assumes that the decision maker is completely unaware of the limitations of their knowledge about the future. However, as surprises are a part of real life, this assumption is too strong and cuts back the power of the theory.

\medskip
\noindent
Both lines of research represent the efforts to include the limitations of a decision maker's knowledge into economic theory. In the remaining parts of this section, we briefly discuss the development of the first line of research mentioned in the introduction which employs alternative concepts of probability measures, and then, in the next section, we turn to review representations of non-knowledge by means of various formalizations of the state space.

\medskip
\noindent
Knight's (1921)~\ct{kni1921} work, and later Ellsberg's (1961)~\ct{ell1961} paradox, gave way to the intuition that there are differences in how people assign and treat probability measures, and that those differences are related to the quality of the decision maker's knowledge. Some probability measures are based on more or less reliable information (evidence, or knowledge), and some result from a default rule based on ignorance. For example, there should be a difference between probability as formed by an expert and by a layman. The intuition behind Schmeidler's (1989)~\ct{sch1989} \emph{non-additive prior probability measures framework} is exactly this: there is a difference between
\begin{quote}
``\ldots bets on two coins, one which was extensively tested and was found to be fair, and another about which nothing is known. The outcome of a toss of the first coin will be assigned a 50--50 distribution due to `hard' evidence. The outcome of a toss of the second coin will be assigned the same distribution in accord with Laplace's principle of indifference. But as Schmeidler (1989) argues, the two distributions feel different, and, as a result, our willingness to bet on them need not to be the same'' (Gilboa \emph{et al} 2008~\ct[p~179]{giletal2008}).
\end{quote}

\medskip
\noindent
The failure of Savage's (1954)~\ct{sav1954} model to account for differences in the knowledge quality in both cases was called by Gilboa \emph{et al} (2008)~\ct[p~181]{giletal2008} ``an agnostic position.'' Ellsberg (1961)~\ct{ell1961} underlined this issue empirically and demonstrated that the preference of a decision maker for ``known'' probabilities violates the ``sure thing principle'' in Savage's axiomatization: people do not necessarily behave as though they were subjective expected utility maximizers. To model a decision maker's state of imperfect knowledge in such situations more accurately, there were suggestions to replace the unique prior probability measure with an entire \emph{set of prior probability measures} (Gilboa and Schmeidler 1989~\ct{gilsch1989}, Bewley 1986, 2002~\ct{bew1986,bew2002}): non-knowledge regarding the likelihood of uncertain states of Nature here is linked to the number of elements contained in a decision maker's set of prior probability measures used in calculations of expected utility of acts and consequences, and so is represented in a more comprehensive fashion than in Savage's (1954)~\ct{sav1954} framework. For example, to account for their ignorance, the decision maker assigns not a unique prior probability to an event but rather a certain \emph{range} of values. In the case of the untested coin (when knowledge of the coin's properties is vague or non-existent) this range for head/tail could be ``between 45 and 55 percent.'' We note that Epstein and Wang (1994)~\ct{epswan1994} later extended Gilboa and Schmeidler's (1989)~\ct{gilsch1989} multiple-priors approach to intertemporal settings.

\medskip
\noindent
A different way to account for the limitations of a decision maker's knowledge of future contingencies was the development of \emph{non-probabilistic concepts}, for example, fuzzy logic and possibility theory (Zadeh 1965, 1978~\ct{zad1965,zad1978}; Dubois and Prade 2011~\ct{dubpra2011}). Interestingly, the economist Shackle (1961)~\ct{sha1961}, whose work was ignored for decades, was one of the founders of this particular line of research. For Shackle, \emph{possibility} in particular expresses the incompleteness of a decision maker's knowledge about the future, and hereby allows representing the ``\emph{degree of potential surprise}'' of an event. Possibility as a measure of subjective non-knowledge is less precise (``fuzzier'') than probability and is based either on a numerical (quantitative) or on a qualitative scaling of events from ``totally possible'' to ``impossible.'' Those measures must not be additive. It means that two or more events can be simultaneously considered as absolutely possible (or impossible, "surprising"). The modern formalized version of this idea suggests that there is a finite set of states to which a possibility distribution is assigned (Dubois and Prade 2011~\ct[p~3]{dubpra2011}):
\begin{quote}
``A possibility distribution is a mapping $\pi$ from $S$ to a totally ordered scale $L$, with top~1 and bottom~0, such as the unit interval. The function $\pi$ represents the \emph{state of knowledge of an agent} (about the actual state of affairs) distinguishing what is plausible from what is less plausible, what is the normal course of things from what is not, what is surprising from what is expected.'' 
\end{quote}

\medskip
\noindent
Despite this seemingly radical innovation, and some promising applications in the economic science (e.g., Dow and Ghosh 2009~\ct{dowgho2009}), the possibility theory could not ``revolutionize'' decision theory. Zadeh (1978)~\ct[p~7]{zad1978}, the founder of fuzzy logic, famously hinted that ``our intuition concerning the behaviour of possibilities is not very reliable'' and required the axiomatization of possibilities ``in the spirit of axiomatic approaches to the definition of subjective probabilities,'' i.e., in line with Savage's (1954)~\ct{sav1954} axiomatization. To make the connection with decision theory, such a theoretical framework was successfully developed by Dubois \emph{et al} (2001)~\ct{dubetal2001}. Also, more generally, various probability--possibility transformations were discussed, i.e., how to translate for example quantitative possibilities into probabilities and vice versa. In the end, as Halpern (2005)~\ct[p~40]{hal2005} states, ``possibility measures are yet another approach to assigning numbers to sets,'' implying all benefits and limits of alternative probability theories.

\medskip
\noindent
Finally, --- and this is very crucial for our discussion --- note that the set of possible consequence-relevant states of Nature in all cases discussed in this section, i.e., in the case of a unique prior probability measure, in the case of a set of prior probability measures, for non-additive prior probability measures, and in the possibility framework, is assumed to be finite, so that a real surprise (a completely unexpected event) cannot be incorporated. However, to properly account for surprising events, this list should not be modelled as exhaustive. It is crucial to emphasise in this context that assigning subjective probability zero does not help to represent true unawareness of particular events because
\begin{quote}
``[s]tatements like `I am assigning probability zero to the event $E$ because I am unaware of it' are nonsensical, since the very statement implies that I think about the event $E$'' (Schipper 2013~\ct[p~739]{sch2013}; cf. also Dekel \emph{et al} 1998~\ct{deketal1998}).
\end{quote}

\medskip
\noindent
By definition, a decision maker should be perfectly unaware of surprising events before committing to a specific action, and it lies in this very nature that this issue cannot be captured solely by means of more or less well-defined probability measures.

\section[Non-knowledge of the state space and the 
possibility of surprises]{Second way of formalization: 
genuine non-knowledge of the state space and the possibility of 
true surprises}
\lb{sec4}
The second line of thought of incorporating non-knowledge on a decision maker's part into economic theory likewise has its history and tradition. It was recognized by a number of authors that in order to include true non-knowledge concerning future contingencies and surprising events into the framework of decision theory, it is necessary to shift research efforts from the issue of determination of adequate (prior) probability measures (i.e., risk and uncertainty in the modern economic parlance) to the issue of representation of a decision maker's \emph{unawareness} with respect to possible states of Nature beyond their imagination which could also affect the consequences of their choice behaviour. This unawareness may be interpreted as a manifestation of a decision maker's natural bounded rationality. Their non-knowledge should not be limited to just a lack of knowledge as to which state from the exhaustive list of states of Nature will materialize (``uncertainty about the true state''), but rather non-knowledge about the full state space itself should be a part of decision theory. This challenge was met in the economics literature in particular by Kreps (1979)~\ct{kre1979}, Fagin and Halpern (1988)~\ct{faghal1988}, Dekel \emph{et al} (1998, 2001)~\ct{deketal1998,deketal2001}, and Modica and Rustichini (1999)~\ct{modrus1999}. Their proposals presuppose a coarse (imperfect) subjective knowledge of all consequence-relevant states of Nature possible, and so criticize a central assumption in Savage's (1954)~\ct{sav1954} and Anscombe and Aumann's (1963)~\ct{ansaum1963} axiomatizations of a decision maker's choice behaviour, suggesting a radical departure from their frameworks. First of all, proving two famous impossibility results, Dekel, Lipman and Rustichini (1998)~\ct{deketal1998} demonstrated that the standard partitional information structures of economic theory (i.e., the set-theoretic state space models discussed earlier in Section~\ref{sec2}) preclude unawareness. Specifically, in such settings, only two very extreme situations can be captured: either a decision maker has \emph{complete knowledge} of the full space of consequence-relevant states of Nature (as has the modeller), or they have \emph{no knowledge} of this state space whatsoever. In addition, Dekel \emph{et al} (1998)~\ct{deketal1998} made explicit crucial epistemic properties of true unawareness: e.g., that it is necessarily impossible for a decision maker to be aware of their own unawareness (technically termed AU introspection); cf. also Heifetz \emph{et al} (2006)~\ct{heietal2006}.

\medskip
\noindent
Following this discussion, new accounts were developed which suggested different ways to depart from the set-theoretic state space concepts of Savage (1954)~\ct{sav1954} and of Anscombe and Aumann (1963)~\ct{ansaum1963}; foremost from their assumption on the existence of an exhaustive list of mutually exclusive consequence-relevant states of Nature which is available to both the modeller and the decision maker alike. These new accounts formalize a principally different kind of non-knowledge compared to the non-knowledge of (prior) probability measures over a complete state space: \emph{unawareness} of potentially ensuing important events, or of additional future subjective contingencies. In terms of elements of the decision matrix in Figure~\ref{fig:fig1}, \emph{only} the modeller now has complete knowledge of all consequence-relevant states of Nature, and of all possible outcomes/lotteries over outcomes contingent on these states. The decision maker has a restricted perception of matters depending on the awareness level they managed to attain.

\medskip
\noindent
Three ways to overcome Dekel \emph{et al}'s (1998)~\ct{deketal1998} impossibility results concerning standard partitional information structures can be identified in the economics literature, two of which maintain the status of a (now enriched) state space concept as a primitive of the framework proposed. These are
\begin{enumerate}

\item the two-stage choice approach

\item the epistemic approach, and

\item the set-theoretic approach.
\end{enumerate}
We now briefly review these in turn.

\medskip
\noindent
One solution is to formalize an endogenous subjective state space of a decision maker as a \emph{derived concept}, as was initially suggested by Kreps (1979, 1992)~\ct{kre1979,kre1992}, and then further developed by Dekel \emph{et al} (2001)~\ct{deketal2001} and Epstein \emph{et al} (2007)~\ct{epsetal2007}. These researchers proposed a decision maker who is unaware of certain future subjective contingencies, and a modeller who can infer a decision maker's subjective state space regarding these contingencies from observing the decision maker's choice behaviour. (To a certain extent this strategy can be viewed as analogous to Savage's (1954)~\ct{sav1954} reconstruction of a decision maker's beliefs from their revealed preferences.) Kreps (1979)~\ct{kre1979} developed a \emph{two-stage model} in which a decision maker first chooses from a set of finite action menus. Subsequently, a particular state of Nature is realized. The decision maker chooses a specific action from the selected menu only afterwards. The central idea is that although the decision maker does not know all the states that are possible, they know their subjective subset of possibilities, and this subset is not exogenous. The decision maker anticipates future scenarios which affect their expected later choices from the action menus and their ex ante utility evaluation of these menus. Thus, these scenarios (or the subjective state space) form the basis for ordinal binary preference relations with respect to the menus and can be revealed through observation of those preferences.

\medskip
\noindent
The more unaware a decision maker is regarding consequence-relevant states of Nature, the more flexibility they prefer by choosing the menus during the first phase. This intuition was more rigorously formalized by Dekel \emph{et al} (2001)~\ct{deketal2001}, who provided conditions required to determine the endogenous subjective state space uniquely. For example, they replaced the action menus by menus of lotteries over finite sets of actions, in the spirit of Anscombe and Aumann (1963)~\ct{ansaum1963}. Epstein \emph{et al} (2007)~\ct{epsetal2007} proposed ways for the two-stage choice approach to account for a decision maker's manifested uncertainty aversion according to Ellsberg's (1961)~\ct{ell1961} empirical result. The pioneers of the unawareness concept depart from Savage's (1954)~\ct{sav1954} and Anscombe and Aumann's (1963)~\ct{ansaum1963} axiomatizations by replacing the state space in the list of primitives by a set of menus over actions which are the objects of choice. This theoretical move allows for dealing with unforeseen contingencies due to a decision maker's natural bounded rationality, the latter of which is manifested by their inability to list all the states of the exogenous world that could be relevant. For further details on this approach refer also to Svetlova and van Elst (2012)~\ct{svehve2013}.

\medskip
\noindent
In the \emph{epistemic approach} to formalizing a decision maker's unawareness, initiated by Fagin and Halpern (1988)~\ct{faghal1988}, and subsequently pursued by Modica and Rustichini (1999)~\ct{modrus1999}, Heifetz, Meier and Schipper (2008)~\ct{heietal2008}, and Halpern and R\^{e}go (2008)~\ct{halreg2008}, a modal logic syntax is employed to elucidate the fine-structure of the (consequence-relevant) states of Nature. Such states are here perceived as maximally consistent sets of propositions which are constructed from a set of countably many primitive propositions, their binary truth values, and a set of related inference rules defined on the set of propositions. The propositional logic models so obtained extend the standard Kripke (1963)~\ct{kri1963} information structures of mathematics. The concrete awareness level attributed to a decision maker is associated with a specific subset of consistent propositions and their corresponding binary truth values; the awareness level varies with the number of elements in these subsets. Depending on the approach taken, the awareness level of a decision maker in a given state of Nature is expressed in terms of an explicit awareness modal operator defined over propositions (Fagin and Halpern 1988~\ct{faghal1988}), or indirectly in terms of a knowledge modal operator (Modica and Rustichini 1999~\ct{modrus1999}, and Heifetz, Meier and Schipper 2008~\ct{heietal2008}).

\medskip
\noindent
While Fagin and Halpern (1988)~\ct{faghal1988} in their multi-person awareness structure deal with a single state space and propose two kinds of knowledge (implicit and explicit) a decision maker may have depending in their awareness level, Modica and Rustichini (1999)~\ct{modrus1999} in their one-person generalized partitional information structure distinguish between the full state space associated with the modeller on the one hand, and the (typically lower-dimensional) subjective state space of the decision maker on the other. A projection operator between these two kinds of spaces is defined. A consequence of this construction is that a 2-valued propositional logic obtains in the full state space, while a 3-valued propositional logic applies in the decision maker's subjective state space: a proposition of which they are not aware at a given state can be neither true nor false. Thus, unawareness of a decision maker of a particular event is given when this event cannot be described in terms of states in their subjective state space. According to Halpern and R\^{e}go (2008)~\ct{halreg2008}, an advantage of addressing the issue of the fine-structure of the states of Nature is that this offers a language of concepts for decision makers at a given state, as well as flexibility for covering different notions of awareness. Furthermore, these authors demonstrated that all of the propositional logic models of the epistemic approach to unawareness referred to above are largely equivalent. So far, propositional logic models have not been tied to any specific decision-theoretic framework.

\medskip
\noindent
The \emph{set-theoretic approach}, finally, can be viewed as a less refined subcase of the propositional logic models of the epistemic approach in that it discards the fine-structure of the states of Nature, thus leading to a syntax-free formalization of unawareness. The key realization here is that in order to overcome Dekel \emph{et al}'s (1998)~\ct{deketal1998} troubling impossibility results regarding a non-trivial representation of unawareness in standard partitional information structures, an entire hierarchy of disjoint state spaces of differing dimensionality should be introduced amongst the primitives of a decision-theoretic framework to describe decision makers that have attained different levels of awareness.

\medskip
\noindent
Heifetz, Meier and Schipper (2006)~\ct{heietal2006} deal with this insight by devising in a multi-person context a finite lattice of disjoint finite state spaces which encode decision makers' different strengths of expressive power through the cardinality of these spaces. Hence, these state spaces share a natural partial rank order between them; every one of them is associated with a specific awareness level of a decision maker. The uppermost state space in the hierarchy of this unawareness structure corresponds to a full description of consequence-relevant states of Nature and may be identified either with an omniscient decision maker or with a modeller. The different state spaces are linked by projection operators from higher ranked spaces to lower ranked spaces. These projection operators are invertible and filter out knowledge existing at a higher level of awareness that cannot be expressed at a lower level. In this way, it is possible to formulate events at a given state of which a decision maker of a certain awareness level has no conception at all. A 3-valued logic applies in each state space, with the exception of the uppermost one where the standard 2-valued logic obtains. Unawareness respectively awareness of a decision maker of a particular event are formally defined indirectly in terms of a knowledge operator, which satisfies all the properties demanded of such an operator in standard partitional information structures; cf. Dekel \emph{et al} (1998)~\ct[164f]{deketal1998}. We remark that the Heifetz \emph{et al} (2006)~\ct{heietal2006} proposal may have the potential to provide a framework for capturing Taleb's (2007)~\ct{tal2007} ``black swan events'' in a decision-theoretic context. For this purpose a scenario is required where no decision maker's awareness level corresponds to the uppermost state space in the hierarchy.

\medskip
\noindent
A related set-theoretic framework was suggested by Li (2009)~\ct{li2009}. In her ``product model of unawareness,'' she distinguishes factual information on the (consequence-relevant) states of Nature from awareness information characterizing a decision maker, and so provides a formal basis for, again, differentiating between the full space of states of Nature and a decision maker's (generically lower-dimensional) subjective state space. With a projection operator between these two spaces defined, events of which a decision maker is unaware can be made explicit.

\medskip
\noindent
In contrast to the epistemic approach, direct contact with decision theory was recently established by Schipper (2013)~\ct{sch2013} for the set-theoretic unawareness structure of Heifetz \emph{et al} (2006)~\ct{heietal2006}. He puts forward an awareness-dependent subjective expected utility proposal in the tradition of Anscombe and Aumann (1963)~\ct{ansaum1963}, where a set of awareness-dependent ordinal binary preference relations for a collection of decision makers is defined over a set of acts on the union of all state spaces in the lattice. Acts map consequence-relevant states of Nature in this union to von Neumann--Morgenstern (1944)~\ct{neumor1944} lotteries over outcomes contingent on these states. That is, preferences for acts can now depend on the awareness-level of a decision maker, and thus may change upon receiving new consequence-relevant information. This is clearly a major conceptual step forward concerning representations of non-knowledge in economic theory, especially since it focuses on the important multi-person case. However, also Schipper's (2013)~\ct{sch2013} proposal is likely to suffer from Ellsberg's (1961)~\ct{ell1961} paradox, as decision makers' experimentally manifested uncertainty aversion has not been formally addressed in his framework. In this respect, one expects that Schipper's (2013)~\ct{sch2013} work could be combined with the multiple-priors methodology of Gilboa and Schmeidler (1989)~\ct{gilsch1989} in order to settle this matter, in analogy to Epstein \emph{et al}'s (2007)~\ct{epsetal2007} extension of the work by Dekel \emph{et al} (2001)~\ct{deketal2001} in the context of the two-stage choice approach.

\section{Discussion}
\lb{sec5}
Having reviewed the major approaches to non-knowledge in static decision-making frameworks, we would now like to address some open questions. We discussed concepts that investigated two key elements of the decision matrix in Figure~\ref{fig:fig1}: the space of consequence-relevant states of Nature, and prior probability measures over this space. Until recently, both approaches have been developed detached from one another: the respective papers have been concerned with \emph{either} the determination of adequate probability measures, \emph{or} with handling imperfect knowledge of the state space. However, the paper by Schipper (2013)~\ct{sch2013} makes an important attempt to connect both of these issues. In our view, further work should be done in this direction.

\medskip
\noindent
Moreover, other elements of the decision matrix, particularly \emph{the set of available actions} and \emph{the set of possible consequences}, have been widely excluded from the discussion about non-knowledge in economics to date. We suggest that more conceptual work should be done to clarify if it is justified to presuppose that actions and their consequences are perfectly known to decision makers, as has been the case in economic decision-making theory so far. Another important open question is how the elements of the decision matrix --- probability measure, state space, actions and consequences --- are related to each other. For example, recent research on performativity, reflexivity and non-linearity (cf. see the recent special issue on reflexivity and economics of the \emph{Journal of Economic Methodology}) suggests that actions chosen could causally influence states of Nature; cf. also Gilboa (2009)~\ct{gil2009}.

\medskip
\noindent
These considerations raise the issue of the very nature of, respectively, the states possible, the actions available, and their resultant consequences. It is important to properly understand what it means \emph{to know} states, actions and consequences, \emph{to be aware or unaware} of them. For example, obviously it makes a difference to conceive of possible states as ``states of nature'' or as ``states of the world'' (Schipper 2013~\ct[p~741]{sch2013}). Both types of states differ concerning the role of the decision maker. In a ``states-of-nature'' approach, the decision maker, i.e., their beliefs and actions, is irrelevant for the construction of the state space: only Nature plays against them. Thus, the elimination of non-knowledge \emph{(of the future?)} would depend on the improvement of our understanding of the physical world. If, however, we conceive of the states as ``states of the world,'' the decision maker's beliefs and actions are a part of the world description, with the necessity to consider the interrelation of all elements of the decision matrix, as well as the interconnections between the decision matrices of different decision makers as a consequence. For the conception of non-knowledge, our understanding of the social world would be as relevant as our views about the physical world. We think these insights, which relate to epistemic game theory, should be further developed, though the complexity of a resultant theoretical framework might become its own constraint.

\medskip
\noindent
Finally, we would like to ask if the assumption of omniscience on the part of the \emph{modeller} in the unawareness concepts reviewed is justified. Is it warranted to presuppose that there is an institution that possesses a complete view of all states of Nature possible, while an ordinary \emph{decision maker} has only imperfect knowledge of them? Heifetz \emph{et al} (2006)~\ct[p~90]{heietal2006} stress that ``\ldots unawareness \ldots has to do with the lack of conception.'' For us, this conception includes knowledge of the interrelationships between all elements of the decision matrix. But who possesses this knowledge? And, given the complexity of those interrelations, \emph{can} anybody possess this knowledge at all? To date, present-day economic modellers have not settled this issue.

\section*{Acknowledgments}
We thank the editors of this volume and an unknown referee for 
useful comments on an earlier draft.

\addcontentsline{toc}{section}{References}


\end{document}